\documentclass[%
reprint,
superscriptaddress,
showpacs,
preprintnumbers,
 bibnotes,
 amsmath,amssymb,
 aps,
prl
]{revtex4-1}

\usepackage{graphicx}
\usepackage{dcolumn}
\usepackage{bm}
\usepackage{color}

\begin{document}


\title{Observation of Unusual Magnetoelastic Effects in a Quasi-1D Spiral Magnet}

\author{Chong~Wang}
\thanks{These authors contributed equally to this study.}
\affiliation{International Center for Quantum Materials, School of Physics, Peking University, Beijing 100871, China}
\author{Daiwei~Yu}
\thanks{These authors contributed equally to this study.}
\affiliation{International Center for Quantum Materials, School of Physics, Peking University, Beijing 100871, China}
\author{Xiaoqiang~Liu}
\thanks{These authors contributed equally to this study.}
\affiliation{International Center for Quantum Materials, School of Physics, Peking University, Beijing 100871, China}
\author{Rongyan~Chen}
\affiliation{International Center for Quantum Materials, School of Physics, Peking University, Beijing 100871, China}
\author{Xinyu~Du}
\altaffiliation[Present address: ]{Beijing Institute of Nanoenergy and Nanosystems, Chinese Academy of Sciences, Beijing 100083, China}
\affiliation{International Center for Quantum Materials, School of Physics, Peking University, Beijing 100871, China}
\author{Biaoyan~Hu}
\affiliation{International Center for Quantum Materials, School of Physics, Peking University, Beijing 100871, China}
\author{Lichen~Wang}
\affiliation{International Center for Quantum Materials, School of Physics, Peking University, Beijing 100871, China}
\author{Kazuki~Iida}
\affiliation{Comprehensive Research Organization for Science and Society (CROSS), Tokai, Ibaraki 319-1106, Japan}
\author{Kazuya~Kamazawa}
\affiliation{Comprehensive Research Organization for Science and Society (CROSS), Tokai, Ibaraki 319-1106, Japan}
\author{Shuichi~Wakimoto}
\affiliation{Quantum Beam Science Center, Japan Atomic Energy Agency, Tokai, Ibaraki 319-1195, Japan}
\author{Ji~Feng}
\email[]{jfeng11@pku.edu.cn}
\affiliation{International Center for Quantum Materials, School of Physics, Peking University, Beijing 100871, China}
\affiliation{Collaborative Innovation Center of Quantum Matter, Beijing 100871, China}
\author{Nanlin~Wang}
\affiliation{International Center for Quantum Materials, School of Physics, Peking University, Beijing 100871, China}
\affiliation{Collaborative Innovation Center of Quantum Matter, Beijing 100871, China}
\author{Yuan~Li}
\email[]{yuan.li@pku.edu.cn}
\affiliation{International Center for Quantum Materials, School of Physics, Peking University, Beijing 100871, China}
\affiliation{Collaborative Innovation Center of Quantum Matter, Beijing 100871, China}


\begin{abstract}
We present a systematic study of spin and lattice dynamics in the quasi-one-dimensional spiral magnet CuBr$_2$, using Raman scattering in conjunction with infrared and neutron spectroscopy. Along with the development of spin correlations upon cooling, we observe a rich set of broad Raman bands at energies that correspond to phonon-dispersion energies near the one-dimensional magnetic wave vector. The low-energy bands further exhibit a distinct intensity maximum at the spiral magnetic ordering temperature. We attribute these unusual observations to two possible underlying mechanisms: (1) formation of hybrid spin-lattice excitations, and/or (2) ``quadrumerization'' of the lattice caused by spin-singlet entanglement in competition with the spiral magnetism.
\end{abstract}

\maketitle


Multiferroic spiral magnets \cite{SpaldinScience2005,CheongNatMater2007,KhomskiiPhys2009,TokuraAdvMater2010} offer a useful test ground for us to gain insight into the coupling between the spin and lattice degrees of freedom. While extensive understanding of magnetoelastic effects have been attained in the static regime \cite{KatsuraPRL2005,SergienkoPRB2006,MostovoyPRL2006,XiangPRL2008,DongPRB2008,LuPRB2015}, investigation of their counterparts in the dynamic regime has proved a more demanding task. The challenge is in part brought about by a rich yet diverse set of experimental observations in both spiral \cite{ValentinePRB2014,PetitPRL2007,ZhangPRB2014,KadlecPRB2014,RovillainNatMater2010,PimenovNatPhys2006,SushkovPRL2007,TakahashiNatPhys2012,KubackaScience2014,SenffPRL2007,RovillainPRL2011} and colinear magnets \cite{GarciaFloresPRL2012,DaiPRB2000,MoussaPRB2003,WagmanPRB2015}, for which a unified theory is still lacking. To make progress in this direction, it is desirable to study materials with simple crystal and magnetic structure, so that the lattice and spin dynamics can be separately determined and compared.

An even more interesting case is when spiral magnetism meets low dimensionality. In reduced dimensions, long-range magnetic order becomes unstable against thermal and/or quantum fluctuations, whereas local entanglement of spins (\textit{i.e.}, spin singlets) becomes more favorable since each spin has only a small number of interacting neighbors. Competition between N\'{e}el-type long-range magnetic order and spin-singlet formation has been widely explored in one-dimensional (1D) antiferromagnetic chains, with in-depth investigations both in theory \cite{Bethe1931,Lieb1961,Haldane1983PRL,Affleck1989} and in experiments particularly for the case of spin-$\frac{1}{2}$ systems \cite{CuGeO3OseroffPRL1995,KCuF3Lake2005,BaCu2Si2O7TsukadaPRB1999,*KenzelmannPRB2001}. Low-dimensional spiral magnets, which commonly host frustrating spin interactions, are particularly interesting because magnetic frustration may further promote spin-singlet formation \cite{HaldanePRB1982,CastillaPRL1995,BuechnerPRL1996}. As spin-singlet valence bonds and lattice dimerization are often two sides of the same coin in real materials \cite{CrossPRB1979,WelleinPRL1998,BeccaPRL2003}, this provides a second route to magnetoelastic coupling, distinct from the one related to spiral magnetism which requires explicit consideration of spin-orbit interactions \cite{CheongNatMater2007,KhomskiiPhys2009,TokuraAdvMater2010}.

The recently discovered multiferroic material CuBr$_2$ \cite{ZhaoAdvMater2012} presents an interesting case in this regard. CuBr$_2$ has a simple crystal structure that belongs to the monoclinic space group $C12/m1$ (\#12), with only three atoms in the primitive cell. The structure consists of edge-sharing CuBr4 squares that form ribbons running along the $b$ axis. Each ribbon constitutes a spin-$\frac{1}{2}$ chain with dominating next-nearest-neighbor antiferromagnetic spin interactions, whereas the nearest-neighbor (ferromagnetic) and inter-chain spin interactions are considerably weaker \cite{LeePRB2012}, rendering the system as quasi-1D. Because of the frustrating intra-chain interactions and the presence of  inter-chain interactions, an incommensurate spiral magnetic order develops below $T_\mathrm{N}$ = 73.5 K with a propagating wave vector $\textbf{Q}_\mathrm{AF}$  = (1, 0.235, 0.5) in reciprocal lattice units (r.l.u.) \cite{ZhaoAdvMater2012,LeePRB2012,Lebernegg_2013_PRB}. The component $q_\mathrm{M} = 0.235$ along the $\hat{\mathbf{b}}^*$ direction corresponds to about 85$^\circ$ spin rotation between adjacent Cu along the chain. A sketch of the crystal and spin structure can be found in Supplemental Material (SM). Such a spin pattern breaks the inversion symmetry and gives rise to spontaneous ferroelectric polarization below $T_\mathrm{N}$ via the inverse-Dzyaloshinskii-Moriya mechanism \cite{SekiPRB2010}. Here, we report a systematic characterization of dynamic signatures of magnetoelastic coupling in CuBr$_2$ that are likely related to the spiral magnetism and/or the low dimensionality of the system.

\begin{figure}
\includegraphics[width=3.300in]{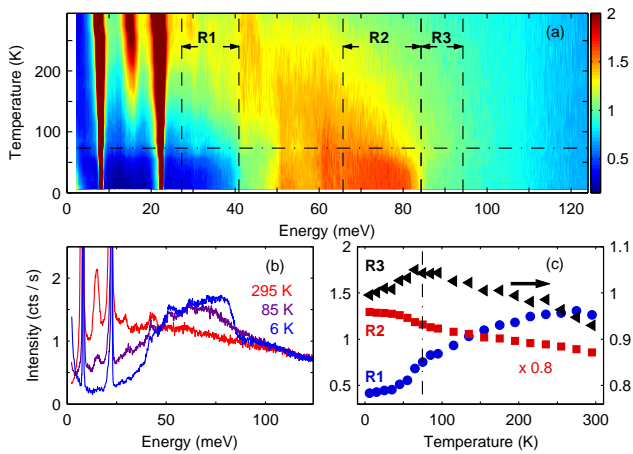}
\caption{\label{fig:one}
(a and b) Variable-$T$ Raman spectra obtained in the $aa$ geometry. The peak at about 15 meV (between the two sharp phonon peaks) originates from two-phonon scattering. Phonon signals below 30 meV deeply saturate the color scale in (a) which is chosen to highlight the high-energy features.  R1-R3 in (a) denote three representative spectral ranges, the intensities averaged over which are displayed in (c). Dashed-dotted lines indicate $T_\mathrm{N}$.
}
\end{figure}

Throughout our presentation, the polarization geometries of infrared (Raman) experiments are indicated by one (two) italic letter that specifies the incoming (incoming and scattered) photon polarization with respect to crystallographic directions. A detailed description of our experimental methods can be found in SM. Figure 1(a-b) displays Raman spectra obtained in the $aa$ geometry over a wide temperature ($T$) and energy range. Upon cooling, a broad signal develops with an increasing characteristic energy, and intensities averaged over three representative spectral ranges (R1-R3), which are calculated by integrating the areas,  all show clear anomalies at $T_\mathrm{N}$ [Fig.~\ref{fig:one}(c)]. The $T$ dependence, together with the distribution of spectral weight primarily in the 40-100 meV range at low temperatures in accordance with estimated strength of spin interactions \cite{LeePRB2012}, indicates that the signal originates from spin excitations and is presumably dominated by two-magnon scattering \cite{FleuryPR1968}. Although becoming very broad, the signal persists to temperatures well above $T_\mathrm{N}$, suggesting that short-range spin correlations are present even at room temperature. The spectral weight transfer from low to high energy below $T_N$ indicates the development of a spin gap, consistent with our neutron scattering results in Fig.~3(a-b).

There are a total of six optical phonon branches in CuBr$_2$. At the Brillouin zone (BZ) center, three modes are Raman-active ($2\times A_g + B_g$) and the remaining three are infrared-active ($A_u + 2 \times B_u$). They can be detected in $aa$- (or $bb$-) and $ab$-polarized Raman spectra, and in $b$- and $a$-polarized infrared spectra, respectively. Indeed, using Raman and infrared measurements, we are able to detect all of them (see SM, Fig.~S3). Moreover, the energies determined form the measurements agree well with the values from the first-principle calculations (see SM for details). Thus, we can be assured of our exhaustive determination of the BZ-center phonons.

\begin{figure}
\includegraphics[width=3.300in]{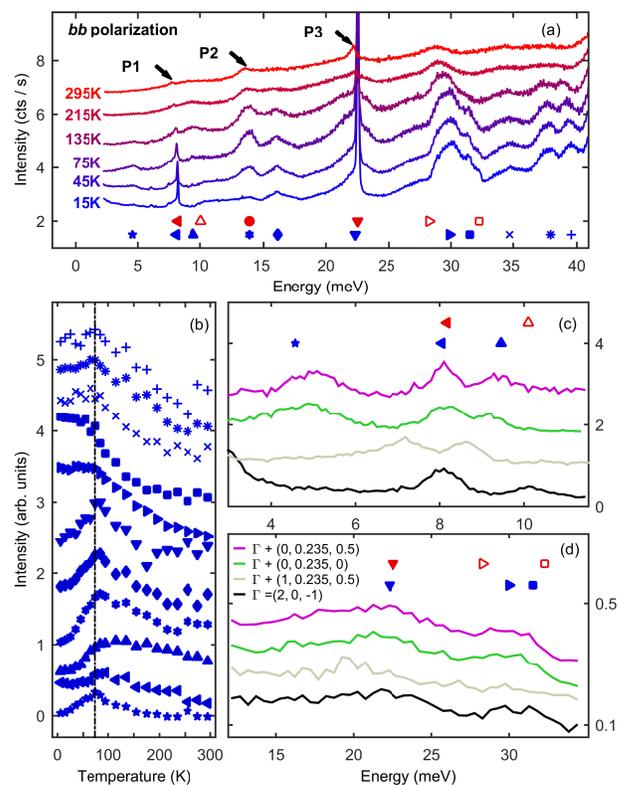}
\caption{\label{fig:two}
(a) Raman spectra obtained in the $bb$ geometry at selected temperatures, offset for clarity. Symbols at the bottom indicate Raman-active optical phonons (red filled), infrared-active optical phonons (red empty) and Raman broad bands (blue), and are coded with data and labels in panels (b-d). (b) $T$ dependence of integrated intensities of broad Raman bands, offset for clarity.  The intensities are determined by fitting the spectra over a nearby energy range to one or two (broad + sharp) peaks on a linear background. (c and d) INS data from energy cuts at four momentum positions, measured at 15 K with incident neutron energies 17 meV (c) and 55 meV (d), offset for clarity. The $\mathbf{b}$-polarized $A_u$ phonon (right empty triangle) is expected not to be observable by INS at the measured $\Gamma$ point (2, 0, -1).
}
\end{figure}

We present our main observation, as seen in Raman spectra obtained in the $bb$ geometry, in Fig.~\ref{fig:two}(a). This geometry is equivalent to $aa$ as far as symmetry-related selection rules are concerned. The spectra are nevertheless very different from those in Fig.~\ref{fig:one}(b) because of difference in the scattering matrix elements. As temperature is lowered from 295 K, we observe a continuous development of a rich set of broad bands: The broad band at P2 has a characteristic energy that is nearly the same as the $B_g$ phonon, but it is not to be mistaken with the phonon which is much sharper in energy (SM, Fig.~S3). Similarly, the two sharp $A_g$ phonon peaks reside on top of broad bands at P1 and P3, but they have very different $T$ dependence of the intensities (SM, Fig.~S4). The combined features at P1 and P3 have an asymmetric Fano line shape \cite{Fano1961} on the phonon peak, indicating possible interference between two scattering processes (SM, Fig.~S5).

In order to attain a comprehensive view of the characteristic energies, the optical phonons and broad bands are labeled by different symbols at the bottom of Fig.~\ref{fig:two}(a). We find that each of the six optical phonons is accompanied by a broad band, except for the $B_g$ mode at 14 meV which is close to two broad bands. Moreover, all broad bands exhibit an intensity anomaly near $T_\mathrm{N}$ [Fig.~\ref{fig:two}(b)]; the fact that many of them are readily observable at high temperatures is in accordance with the presence of short-range spin correlations well above $T_\mathrm{N}$. These results suggest that the broad bands have an origin related to both phonons and magnetism. Broad bands at energies above 33 meV can be related to two-phonon excitations and are thus compatible with this interpretation.

\begin{figure}
\includegraphics[width=3.300in]{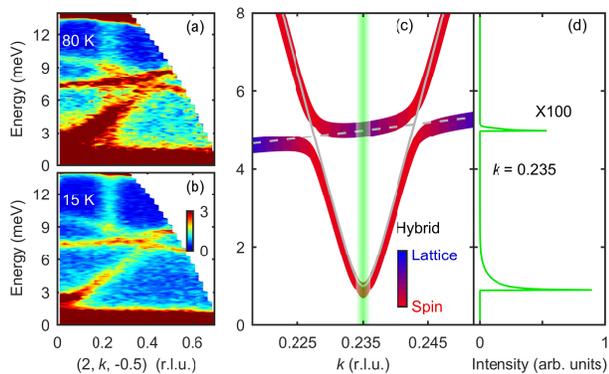}
\caption{\label{fig:three}
(a and b) Phonon and magnon dispersions determined by INS at 80 K and 15 K, respectively.
(c) Schematic of hybridization between phonons (grey dashed line) and magnons (grey solid line), which leads to opening of hybridization gaps and a redistribution of spin and lattice contribution to the eigenvectors.
The Raman spectrum in (d) is calculated based on the new momentum selection rule at the magnetic wave vector (green vertical line) using the schematic magnon-phonon dispersion in (c) (see SM for details).
The bottom of the magnon dispersion is too low in energy to be observed in our Raman experiment away from the elastic line.
}
\end{figure}

To understand why the broad-band energies are close but \textit{not} exactly equal to the phonon energies at the BZ center, we resort to a comparison with our inelastic neutron scattering (INS) experiment, which allows us to detect phonons away from the BZ center. Data of several energy cuts, obtained at 15 K, are displayed in Fig.~\ref{fig:two}(c-d), together with phonon and broad-band labels after those in Fig.~\ref{fig:two}(a). At the $\Gamma$ point, we find an excellent agreement among our Raman, infrared, and INS determination of the phonon energies, but there is no INS signal that corresponds to the broad bands at 4.7 and 9.4 meV [Fig.~\ref{fig:two}(c)]. Instead, INS phonon peaks, with continuous dispersion throughout the BZ, are found at these energies at momentum positions that are offset from $\Gamma$ by $q_\mathrm{M}$ along $\mathbf{b^*}$, or further by 0.5 r.l.u. along $\mathbf{c^*}$ (and anywhere in between). The weak dispersion along $\mathbf{c^*}$ is due to the weak inter-layer van der Waals interactions \cite{ZhaoAdvMater2012}. A particularly revealing case is the broad band at about 4.7 meV, which is below all optical phonon branches. Figure~\ref{fig:three}(a) shows, in another equivalent Brillouin zone, that the corresponding INS peak is on the dispersion of an acoustic phonon branch. We therefore conclude that the broad bands are connected to phonon dispersions at $q_\mathrm{M} \hat{\mathbf{b}}^*$.

Here we discuss two possible scenarios that may explain our observation. In the first scenario, since the broad-band energies are related to phonon dispersions near $q_\mathrm{M} \hat{\mathbf{b}}^*$, they may result from finite-momentum bosons (\textit{i.e.}, phonon and magnon) in the presence of quasi-static spin correlations. We outline here the conceptual thrusts for the new Raman momentum selection rule, whereas the theoretical derivations that lead to a calculable model is detailed in the SM. The conventional Raman scattering process for phonons involves three steps: (1) A photon is absorbed and the material makes a transition to a virtual electronic excited state. (2) Electron-phonon interaction causes energy transfer to the lattice in the form of a zone-center phonon. (3) The electronic system relaxes and a photon with less energy (Stokes scattering) is emitted. With spin-orbit coupling, however, the second step can take an alternative route, leading to intermediate states with magnon excitations \cite{FleuryPR1968,ShenPR1966}. In the presence of spin correlations characterized by wave vector $\mathbf{q}_\mathrm{s}$, creation and annihilation of linear magnons near $\mathbf{q}_\mathrm{s}$ becomes possible. The excitations of magnons alone, however, cannot produce distinct peak-like structures in a Raman spectrum.  It must be aided by magnon-phonon hybridization, which can lead to a situation illustrated in Fig.~\ref{fig:three}(c) calculated based on a schematic hybridization model (see SM for details). The hybridization results in a non-zero contribution from spins on the ``phonon'' branch. In the 1D limit, this gives rise to van Hove singularities so that the hybridized modes produce a peak in the Raman spectrum [Fig.~\ref{fig:three}(d)]. Just above $T_\mathrm{N}$, the inter-chain ordering is lost and the spin correlations become mostly 1D, as is confirmed by the collapse of spin gaps at the 1D but not the 3D magnetic wave vector [Fig.~\ref{fig:three}(a-b)]. The Raman peak is hence expected to be maximized at $T_\mathrm{N}$, consistent with our experimental observation. It must be emphasized that the present theory in this scenario requires both spin-orbit coupling and the finite $\mathbf{q}_\mathrm{s}$ spin correlation at the same time. Without spin-orbit coupling, the Raman process cannot involve magnon excitations. Without the spin-correlation at $\mathbf{q}_\mathrm{s}$, the Raman process can only involve zone-center bosons, without the new momentum selection rule.

\begin{figure}
\includegraphics[width=3.300in]{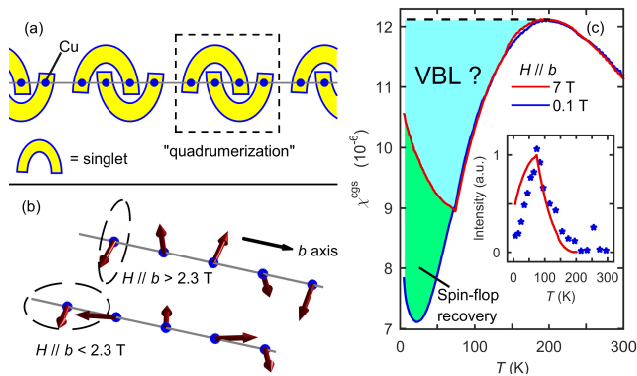}
\caption{\label{fig:four}
(a) Schematic of possible spin-singlet formation in a Cu chain. Each singlet occupies a pair of next-nearest neighbors, causing the affected Cu$^{2+}$ ions to slightly approach each other. The lattice is therefore deformed in a ``quadrumerized'' fashion. (b) Schematic of spin-flop transition below and above a critical magnetic field of $\approx2.3$ Tesla applied along the $b$ axis. (c) Uniform magnetic susceptibility of CuBr$_2$ measured in low and high magnetic fields. The inset displays a comparison of the $T$-dependent intensity of the Raman broad band near 4.7 meV (Fig.~\ref{fig:two}) and the depletion of uniform magnetic susceptibility below 200 K measured in a magnetic field of 7 Tesla along the $b$ direction.
}
\end{figure}

In our second scenario, the Raman broad bands arise from regular phonons near the 1D wave vector $q_\mathrm{M} \hat{\mathbf{b}}^*$, which become back-folded to the nominal BZ center in the presence of quasi-static lattice distortions. However, the long-range spiral magnetic order is not expected to cause lattice distortions with wave vector $q_\mathrm{M} \hat{\mathbf{b}}^*$ (but instead, at $q=0$ and/or $2q_\mathrm{M} \hat{\mathbf{b}}^*$), and an origin related to the spiral magnetism is further incompatible with the decrease of Raman intensities below $T_\mathrm{N}$. To overcome this difficulty, we look into the possibility of alternative spin correlations. In the limit that only the dominant antiferromagnetic interactions between next-nearest neighbors are present, the spin system of CuBr$_2$ becomes fully 1D and each CuBr2 ribbon can be viewed as two interpenetrating spin-1/2 antiferromagnetic chains.  With the help of a deformable lattice, it has been demonstrated in various spin-$\frac{1}{2}$ antiferromagnetic chain compounds \cite{NaV2O5Isobe1996,TTFJacobsPRB1976,CuGeO3HasePRL1993} that a spin-Peierls state \cite{CrossPRB1979} can be stabilized at low temperatures, which breaks the lattice translational symmetry by forming a crystalline arrangement of spin-singlet valence bonds. Indeed, back-folding of phonons has been observed with Raman scattering in spin-Peierls compounds both in the spin-Peierls state \cite{CuGeO3KuroePRB1994,Kuroe1998} and in the short-range ordered state \cite{CuGeO3Loosdrecht1996PRL}. Figure~\ref{fig:four}(a) illustrates the possible situation in a Cu chain of CuBr$_2$. Upon the putative formation of next-nearest-neighbor spin singlets, the Cu chain will ``quadrumerize'' with wave vector $0.25 \hat{\mathbf{b}}^*$. This wave vector is indistinguishable from $q_\mathrm{M} \hat{\mathbf{b}}^*$ concerning phonon-dispersion energies, so it will be consistent with our observation. Since no transition to a spin-Peierls state has been identified in CuBr$_2$, and because our Raman features are broad, we think it is possible that CuBr$_2$ is in a ``valence-bond liquid'' (VBL) state above $T_\mathrm{N}$. The competition between spin-singlet formation and long-range spiral magnetic order can then explain the unusual $T$ dependence of the Raman intensities.

A further piece of experimental evidence in support of the second scenario is presented in Fig.~\ref{fig:four}(b-c). In the long-range spiral magnetic state, CuBr$_2$ exhibit easy-plane spin anisotropy with the $b$ axis lying in the easy plane \cite{ZhaoAdvMater2012,BanksPRB2009}. A spin-flop transition can hence occur if a sufficiently large magnetic field is applied along $b$, as is indeed observed above $H_\mathrm{SF}\approx$ 2.3 T (SM, Fig.~S6). In a magnetic susceptibility measurement, a large part of the susceptibility depletion due to the formation of spin spirals will thus be recovered \cite{PbCuSO4PRBschapers2013} if the measurement is performed in a magnetic field greater than $H_\mathrm{SF}$. Our measurements performed in fields of 0.1 T and 7 T [Fig.~\ref{fig:four}(c)] confirm that this is at least partially the case below $T_\mathrm{N}$. However, the susceptibility depletion starts already below $\sim200$ K which is well above $T_\mathrm{N}$. The fact that no spin-flop recovery of the susceptibility can be observed between $T_\mathrm{N}$ and 200 K suggests that the depletion is not caused by spiral spin correlations, but probably by the presence of a VBL state. In fact, we do not observe a full spin-flop recovery of the susceptibility even below $T_\mathrm{N}$,  implying that the relevance of the assumed VBL state persists even deeply into the magnetically ordered phase. Remarkably, the depleted magnetic susceptibility not recovered in the high-field measurement exhibits a temperature dependence very similar to that of the Raman broad-band intensities [Fig.~\ref{fig:four}(c) inset]. If the VBL scenario is true, CuBr$_2$ presents a dimensional crossover from 1D (well above $T_\mathrm{N}$) to 3D physics (near and below $T_\mathrm{N}$), and despite the 3D long-range order eventually wins, short-range entanglement of spins and a locally quadrumerized lattice exist at all times. To our knowledge, such a crossover has never been observed in a spiral magnet.

To summarize, we have reported spectroscopic evidence for magnetoelastic coupling in CuBr$_2$. The phenomena are consistent with phonon back-folding from near the quasi-1D spiral magnetic wave vector, although the temperature dependence of the Raman intensities requires additional thoughts. We attribute our observation to the formation of hybrid spin-lattice excitations near the spiral magnetic wave vector, and/or to the short-range formation of spin singlets with local lattice deformations in competition with the spiral magnetism. The quasi-static lattice deformations in the second scenario are expected to give rise to diffuse signals in $x$-ray scattering experiments, which are currently underway.

\begin{acknowledgments}

We wish to thank P. Bourges, P. Abbamonte, T. Dong, C. Fang, B. Keimer, D.-H. Lee, J. Park, L.-P. Regnault, F. Wang, and W.-Q. Yu for stimulating discussions. Work at Peking University is supported by NSFC (Nos. 11374024, 11522429, and 11174009) and MOST (Nos. 2013CB921901, 2013CB921903, and 2015CB921302). The neutron experiment at the MLF, J-PARC was performed under a user program (Proposal No. 2014B0032).

\end{acknowledgments}

\bibliography{CuBr2_raman}
\end{document}